\documentclass[]{raa}
\usepackage{cases}            
\usepackage{graphicx,times,subfig}
\usepackage{natbib}
\usepackage{longtable,textcomp}
\usepackage{amssymb}

\begin{document}

   \title{A possible origin of the Galactic Center magnetar SGR 1745-2900}

  \volnopage{Vol.0 (201x) No.0, 000--000}
   \setcounter{page}{1}

   \author{Quan Cheng \inst{1}, Shuang-Nan Zhang \inst{1} and Xiao-Ping Zheng\inst{2}}

\institute{$^1$Key Laboratory of Particle Astrophysics, Institute
of High Energy Physics, Chinese Academy of Sciences, Beijing 100049, China\\
$^2$Institute of Astrophysics, Central China Normal University,
Wuhan 430079, China\\
E-mail address: qcheng@ihep.ac.cn}

\abstract{Since a large population of massive O/B stars and putative
neutron stars (NSs) located in the vicinity of the Galactic center
(GC), intermediate-mass X-ray binaries (IMXBs) constituted by a NS
and a B-type star probably exist here. We investigate the evolutions
of accreting NSs in IMXBs (similar to M82 X-2) with a
$\sim5.2M_\odot$ companion, and orbit period $\simeq2.53$ day. By
adopting an mildly super-Eddington rate
$\dot{M}=6\times10^{-8}M_\odot~{\rm yr}^{-1}$ for the early Case B
Roche-lobe overflow (RLOF) accretion, we find only in accreting NSs
with quite elastic crusts (slippage factor $s=0.05$), the toroidal
magnetic fields can be amplified within 1 Myrs, which is assumed to
be the longest duration of the RLOF. These IMXBs will evolve into
NS+white dwarf (WD) binaries if they are dynamical stable. However,
before the formation of NS+WD binaries, the high stellar density in
the GC will probably lead to frequent encounters between the
NS+evolved star binaries (in post-early Case B mass transfer phase)
and NSs or exchange encounters with other stars, which may produce
single NSs. These NSs will evolve into magnetars when the amplified
poloidal magnetic fields diffuse out to the NS surfaces.
Consequently, our results provide a possible explanation for the
origin of the GC magnetar SGR 1745-2900. Moreover, the accreting NSs
with $s>0.05$ will evolve into millisecond pulsars (MSPs).
Therefore, our model reveals the GC magnetars and MSPs could both
originate from the special kind of IMXBs.
 }

    \authorrunning{Q. Cheng, S. N. Zhang \& X. P. Zheng}            
    \titlerunning{A possible origin of SGR 1745-2900}  
    \maketitle

%
\section{Introduction}           

Magnetars are generally regarded as strongly magnetized neutron
stars (NSs). Release of their magnetic energy could power the burst
activities of soft gamma-ray repeaters and/or anomalous X-ray
emissions of some pulsars (Thompson \& Duncan 1995, 1996). Moreover,
the plateau afterglows of some gamma-ray bursts (GRBs), the light
curves of superluminous supernovae and bright mergernovae can be
well understood in the magnetar scenario, which suggests that the
central objects are newly born, millisecond rotating magnetars with
surface dipole magnetic fields $B_{\rm s}=10^{14-15}$ G (e.g., Dai
\& Lu 1998; Zhang \& M\'esz\'aros 2001; Fan \& Xu 2006; Kasen \&
Bildsten 2010; Yu et al. 2010; Dall'Osso et al. 2011; Rowlinson et
al. 2013; Yu et al. 2013; L\"{u} et al. 2015). Unlike the GRB
magnetars, the Galactic magnetars have relatively long spin periods
(2-12 s) and wide dipole magnetic field distribution, ranging from
$6\times10^{12}$ G to $2\times10^{15}$ G (Olausen \& Kaspi 2014).
Even though SGR 0418+5729 and Swift J1822.3--1606 both have much
lower dipolar fields of $6\times 10^{12}$ G and $1.35\times 10^{13}$
G, respectively (Rea et al. 2013a; Scholz et al. 2014), strong
initial surface dipole and internal toroidal magnetic fields of
$\sim10^{14}$ G and $\sim10^{15-16}$ G are required to recover the
observed properties of the two magnetars (Turolla et al. 2011; Rea
et al. 2012).

To date, 28 magnetars have been discovered: 26 in the Galaxy, one in
the Large Magellanic Cloud, and one in the Small Magellanic
Clouds\footnote{http://www.physics.mcgill.ca/$\sim$pulsar/magnetar/main.html}.
For the Galactic magnetars, more than half of them are located in
the inner Galaxy (Olausen \& Kaspi 2014). The most striking one is
SGR 1745-2900, which has a projected offset of only 0.12 pc from the
Galactic center (GC) black hole, Sgr ${\rm A}^*$ (Rea et al. 2013b).
SGR 1745-2900 was first discovered by \textit{Swift} XRT when it
underwent an X-ray outburst (Kennea et al. 2013). This source was
identified as a magnetar after the discovery of its 3.76 s
pulsations by \textit{NuSTAR} (Mori et al. 2013). Meanwhile, the
spin-down rate of this magnetar is determined to be
$\dot{P}=6.61\times10^{-12}$ s ${\rm s}^{-1}$, which gives a surface
equatorial dipole magnetic field of $1.6\times10^{14}$ G via $B_{\rm
d}=3.2\times10^{19}\sqrt{P\dot{P}}$ (Rea et al. 2013c).
Subsequently, through a 4 months observation using \textit{NuSTAR}
and \textit{Swift} after its discovery, Kaspi et al. (2014) find
that the spin-down rate of SGR 1745-2900 has been increased by about
2.6 times, which yields a stronger dipole field of
$2.3\times10^{14}$ G. Follow-up radio observations also detected
this GC magnetar using the Effelsberg radio telescope, the
Nan\c{c}ay radio telescopes, the Very Large Array, Jodrell Bank, the
Australia Telescope Compact Array, the Green Bank Telescope and the
Parkes Radio Telescope (Eatough et al. 2013; Rea et al. 2013c;
Shannon \& Johnston 2013), making it the fourth radio-magnetar found
in the Galaxy. Most surprisingly, Spitler et al. (2014) suggested
that the pulsed emission from this magnetar could even be detected
at frequencies down to 1.2 GHz, which indicates that the pulse
profile broadening due to the scattering material at the GC may be
much weaker than previously predicted. As a result, if the
scattering of the whole GC region is indeed much weaker, many
canonical and millisecond pulsars (MSPs) should have been detected
by previous surveys (Spitler et al. 2014), since $\sim1000$ radio
pulsars are predicted to orbit Sgr ${\rm A}^*$ with periods less
than 100 yrs (Pfahl \& Loeb 2004). However, no normal pulsars have
been detected within the central parsec of Sgr ${\rm A}^*$ through
several surveys at different observing frequencies (Johnston et al.
1995; Johnston et al. 2006; Deneva et al. 2009; Macquart et al.
2010; Bates et al. 2011).

Many reasons may be responsible for the lack of detection of GC
pulsars: spatially complex scattering (Spitler et al. 2014),
intrinsic deficit of pulsars (Spitler et al. 2014), and MSPs
dominant pulsar population which cannot be found by previous GC
pulsar searches (Macquart \& Kanekar 2015). Specifically, the
intrinsic deficit of pulsars may result from dark matter
accretion-induced collapse of GC NSs (Fuller \& Ott 2015), or
efficient formation of magnetars rather than ordinary pulsars in the
GC as implied by the discovery of SGR 1745-2900 (Dexter \& O'Leary
2014). Since a large amount of massive and strongly magnetized O/B
stars exist in the GC, their collapses could efficiently produce
magnetars, which spin down very rapidly (Dexter \& O'Leary 2014).
Moreover, a large amount of MSPs are suggested to exist in the GC
due to the very dense stellar environment (Spitler et al. 2014;
Dexter \& O'Leary 2014; Macquart \& Kanekar 2015).

Inspired by this suggestion, we propose that the GC is indeed a
region very efficient for magnetar formation. However, these
magnetars could originate from the NSs in intermediate-mass X-ray
binaries (IMXBs) comprised by a NS and a massive B-type
star\footnote{The B-type stars usually have masses in the range
$2.1M_\odot-16M_\odot$ (Habets \& Heintze 1981), thus they have
relatively long main-sequence lifetimes of $10^7-1.6\times10^9$ yrs
(Bhattacharya \& van den Heuvel 1991).}. The predicted large
population of NSs in combination with a large amount of massive
B-type stars will reasonably result in many IMXBs in the GC.
Specifically, we consider the evolutions of accreting NSs in a
special kind of IMXBs, of which the companions are evolved stars
with mass $M_{\rm c}\sim5.2M_\odot$, and orbital period $P_{\rm
orbit}\simeq2.53$ day. In such an IMXB system Roche-lobe overflow
(RLOF) accretion could take place at the end of the main sequence of
the massive companion when it starts to expand (namely Hertzsprung
gap phase) (Karino \& Miller 2016). The early Case B RLOF accretion
proceeds on the thermal timescale with a high accretion rate, which
may be responsible for the high observed X-ray luminosity of the
ultraluminous X-ray source (ULX) M82 X-2 (Bachetti et al. 2014;
Dall'Osso et al. 2015; Karino \& Miller 2016). The RLOF accretion
timescale is some fraction of the time that the companion remains in
the Hertzsprung gap phase. For a $5.2M_\odot$ companion, the
duration of the Hertzsprung gap phase is about 1.5 Myrs (Karino \&
Miller 2016). To accurately determine the timescale of RLOF
accretion, one need to perform a detailed stellar evolution
calculation (Karino \& Miller 2016), which is beyond the scope of
this paper. We assume that the longest duration of RLOF accretion is
1 Myrs. Actually, if the accretion rate is much higher (than the
Eddington limit $\dot{M}_{\rm Edd}$), within 1 Myrs the NS can be
spun up to the critical frequency to trigger the $r$-mode. Thus, a
shorter RLOF accretion timescale will not change our result
qualitatively.

When RLOF accretion commences, in an IMXB of our interest here, the
NS which is initially not in spin equilibrium, will be spun up. The
accreting NS may be observed as an ULX similar to M82 X-2. The high
accretion rate and long enough duration of the RLOF accretion in
such a binary system could easily spin up the NS to millisecond
period if its dipole magnetic field is $10^{8-9}$ G as suggested in
Klu\'zniak \& Lasota (2015). The subsequent evolution of the NS
depends on its properties and the accretion rate. If the spin
equilibrium frequency $\nu_{\rm eq}$ is higher than the critical
frequency $\nu_{\rm cr}$ at which $r$-mode arises, the accreting NS
will succumb to the $r$-mode instability when $\nu_{\rm cr}$ is
reached (Cuofano \& Drago 2010; Cuofano et al. 2012). The
$r$-mode-induced differential rotation could produce strong internal
toroidal magnetic field, which will subsequently succumb to the
Tayler instability, leading to the formation of enhanced poloidal
magnetic field (Cheng \& Yu 2014). Through such a dynamo process,
the accreting NS could evolve into a NS with high internal toroidal
and poloidal magnetic fields. However, if $\nu_{\rm eq}<\nu_{\rm
cr}$, the accreting NS will evolve into a millisecond rotating NS
without enhanced internal magnetic fields since $r$-mode could never
arise.

Evolution of such an IMXB is still an open issue. It is possible
that the delayed dynamical instability will set in after the early
Case B mass transfer due to the large mass ratio between the
companion and the NS (Bhattacharya \& van den Heuvel 1991; Tauris et
al. 2000; Shao \& Li 2012; 2015). The IMXB will experience the
common envelope evolution (CEE), possibly resulting in a merger of
the binary. On the other hand, the CEE is not inevitable even rapid
mass transfer onto the NS occurs as suggested by King \& Ritter
(1999). In this dynamical stable case, the IMXB will finally evolve
into a NS binary made of a NS and a WD. Before the NS+WD system is
finally formed, the NS could still accrete from the evolved
companion after the early Case B mass transfer phase (see e.g.,
Tauris et al. 2000). However, the high stellar density in the
GC\footnote{The stellar density in the GC is much higher than that
in globular cluster cores (see e.g., Macquart \& Kanekar 2015).}
could easily lead to frequent encounters between the NS+evolved star
binary (in post-early Case B mass transfer phase) and NSs or
exchange encounters, which may disrupt the binary system and produce
a single NS (see e.g., Sigurdsson \& Phinney 1995; Ivanova et al.
2008; Verbunt \& Freire 2014). The scenarios were also invoked by
Verbunt \& Freire (2014) to account for the production of isolated
pulsars in some globular clusters. Without further accretion on to
the NS, the enhanced poloidal magnetic field could diffuse out to
the NS surface and change the NS into a magnetar. Hence, when the
magnetar is formed, it should not be in a binary in order to avoid
further large amount of accretion ($\sim0.1M_\odot$ is possible)
(Pan et al. 2013), which may lead to substantial decay of the
magnetic field (Zhang \& Kojima 2006). Our scenario provides another
possible explanation for the origin of the GC magnetar SGR
1745-2900, and suggests that both the GC magnetars and MSPs could
originate from IMXBs.

We note that, in principal, in order to accurately estimate the
evolution of the accreting NS in the IMXB, one should perform
detailed binary evolution calculations. However, we are mainly
concerning with the total amount of angular momentum carried by the
accreted material that would spin up the NS to millisecond period.
Hence, for simplicity, we only consider the early Case B RLOF
accretion phase and assume that the NS accrete material from its
surrounding Keplerian disk at a time-averaged rate $\dot{M}\gtrsim
\dot{M}_{\rm Edd}$. The assumption overestimates the accretion
timescale needed to spin up the NS to millisecond period, because
the accretion rate can actually be extremely super-Eddington as in
the case of M82 X-2 (Klu\'zniak \& Lasota 2015).

Our work is based on the magnetar formation mechanism of Cheng \& Yu
(2014). We suggest the readers to refer to that paper for detailed
formation process of newly born magnetars. This paper is organized
as follows. In Section 2, we present evolutions of $r$-mode, spin,
and magnetic fields of the accreting NSs in the special kind of
IMXBs introduced above. Results are shown in Section 3. Finally,
conclusion and discussions are presented in Section 4.

\section{MODEL}

During the accretion phase, the maximum spin frequency that the
accreting NS could reach is the spin equilibrium frequency $\nu_{\rm
eq}$, which is determined by setting $r_{\rm co}\simeq r_{\rm m}$
with $r_{\rm co}$ and $r_{\rm m}$ representing the co-rotation and
magnetospheric radius, respectively. The $\nu_{\rm eq}$ of an
accreting NS with mass $M=1.4M_\odot$ and radius $R=12.53$ km is
(Alpar et al. 1982)
\begin{eqnarray}
\nu_{\rm eq}\simeq 3.64\times10^2({B_{\rm d}\over10^9~{\rm
G}})^{-6/7}({\dot{M}\over10^{-8}M_\odot~{\rm yr}^{-1}})^{3/7}~{\rm
Hz},\label{veq}
\end{eqnarray}
where $B_{\rm d}$ and $\dot{M}$ are the surface equatorial dipole
magnetic field and time-averaged accretion rate of the NS,
respectively. However, with the spin-up of the accreting NS,
$r$-mode, which arises due to the action of Coriolis force
(Andersson 1998; Friedman \& Morsink 1998), will become unstable if
the star reaches the critical frequency $\nu_{\rm cr}$.
Consequently, the occurrence of $r$-mode instability requires
$\nu_{\rm cr}<\nu_{\rm eq}$. The critical frequency $\nu_{\rm cr}$
is determined via
\begin{eqnarray}
{1\over\tau_{\rm g,r}}-{1\over\tau_{\rm v}}=0,\label{tgtv}
\end{eqnarray}
where $\tau_{\rm g,r}=3.26(\Omega/\sqrt{\pi G\bar{\rho}})^{-6}$ is
the timescale of the $r$-mode-induced GW emission with $\bar{\rho}$
the mean density of the star. $\tau_{\rm v}=(\tau_{\rm
bv}^{-1}+\tau_{\rm sv}^{-1}+\tau_{\rm vbl}^{-1})^{-1}$ is the
viscous damping timescale, where the timescales $\tau_{\rm bv}$,
$\tau_{\rm sv}$, and $\tau_{\rm vbl}$ correspond to the bulk
viscosity, shear viscosity, and dissipation in the viscous boundary
layer at the crust-core interface, respectively. The specific
expressions of $\tau_{\rm bv}$ and $\tau_{\rm sv}$ can be found in
Yu et al. (2009), while $\tau_{\rm vbl}=23.3s^{-2}(\Omega/\sqrt{\pi
G\bar{\rho}})^{-1/2}(T/10^8 {\rm K})$ with $T$ denotes the stellar
internal temperature and $s$ the slippage factor (Lindblom et al.
2000), whose value lies in the range $0.05\lesssim s\lesssim 1$
(Levin \& Ushomirsky 2001). The smaller $s$ is, a more elastic crust
the accreting NS will have. For an accreting NS with typical
temperature $T\sim10^8$ K (Cuofano et al. 2012; Haskell et al.
2012), the viscous damping is dominated by dissipation in the
viscous boundary layer. Hence, following Equation (\ref{tgtv}), the
expression for $\nu_{\rm cr}$ is
\begin{eqnarray}
\nu_{\rm cr}\simeq 936.80\left({T\over10^8~{\rm
K}}\right)^{-2/11}s^{4/11}~{\rm Hz}. \label{vcr}
\end{eqnarray}

$R$-mode in a NS will unavoidably lead to differential rotation,
which cannot only determine the saturation amplitude of the mode
(S\'a \& Tom\'e 2005), but also wind up the initial poloidal
magnetic field to form toroidal magnetic field (Rezzolla et al.
2000). Overall, the growth of $r$-mode can be increased through
gravitational radiation back-reaction; however, in the meanwhile, it
can also be suppressed by viscous damping, and by forming the
toroidal magnetic field. In an accreting NS with a Keplerian disk,
the total stellar angular momentum can be increased by angular
momentum transfer from the accretion material. However, the magnetic
dipole radiation (MDR) and gravitational wave (GW) emissions of the
NS can decrease the total stellar angular momentum. Specifically,
the GW emissions are mainly contributed by $r$-mode and quadrupole
deformation induced by the newly formed strong toroidal magnetic
field. Following Cheng \& Yu (2014) and Cuofano \& Drago (2010), the
evolution equations for the $r$-mode amplitude $\alpha$ and angular
frequency $\Omega$ of the accreting NS can be expressed as
\begin{eqnarray}
{d\alpha\over dt}&=&\left[1+{2\alpha^2\over
15}(\delta+2)\right]{\alpha\over \tau_{\rm g,r}}
-\left[1+{\alpha^2\over 30}(4\delta+5)\right]\left({\alpha\over
\tau_{\rm v}}+{\alpha\over \tau_{\rm
t}}\right)\nonumber\\
&+&{\alpha\over 2\tau_{\rm g,t}}+{\alpha\over 2\tau_{\rm
d}}-{\alpha\dot{M}\over2\tilde{I}\Omega}\left({G\over
MR^3}\right)^{1/2}, \label{alpha}
\\
{d\Omega\over dt}&=&-{4\alpha^2\Omega\over
15}(\delta+2)\left[{1\over \tau_{\rm
g,r}}-{(4\delta+5)\over4(\delta+2)}\left({1\over\tau_{\rm
v}}+{1\over\tau_{\rm t}}\right)\right]\nonumber\\
&-&{\Omega\over \tau_{\rm g,t}}-{\Omega\over \tau_{\rm
d}}-\Omega{\dot{M}\over M}+{\dot{M}\over\tilde{I}}\left({G\over
MR^3}\right)^{1/2}, \label{Omega}
\end{eqnarray}
where $\tilde{I}=0.261$, and $\delta$ denotes the initial amount of
differential rotation. $\tau_{\rm d}={3Ic^3/(2B_{\rm
d}^2R^6\Omega^2)}$ is the MDR timescale with $I=0.261MR^2$
representing the moment of inertia of the NS. Moreover, strong
toroidal magnetic field with volume-averaged strength $\bar{B}_t$
could deform the star into a prolate ellipsoid, whose ellipticity is
$\epsilon=-{{5\bar{B}_{\rm t}^2R^4}/{6GM^2}}$ (Cutler 2002). Hence,
the GW emission timescale of the magnetically-induced deformation is
$\tau_{\rm g,t}=5c^5/(32GI\epsilon^2\Omega^4)$ (Cutler \& Jones
2001). Following Cheng \& Yu (2014), after choosing an initial
internal dipolar magnetic field $B_{\rm ini}=B_{\rm
d}(R/r)^3(2\cos\theta \textbf{e}_r+\sin\theta \textbf{e}_\theta)$,
the growth rate of toroidal magnetic field energy is derived as
\begin{eqnarray}
{dE_{\rm t}\over dt}={75\over 448\pi^2}B_{\rm
d}^2R^3(2\delta+3)^2\alpha^2\Omega\int_0^t\alpha^2\Omega dt'.
\end{eqnarray}
Hence, the toroidal magnetic formation timescale can be written as
(Cheng \& Yu 2014)
\begin{eqnarray}
\tau_{\rm t}=\frac{448\pi^2(4\delta+9)I^*\Omega}{75B_{\rm
d}^2R^3(2\delta+3)^2\int_0^t\alpha^2\Omega dt'},
\end{eqnarray}
where $I^*=1.635\times10^{-2}MR^2$ is an effective momentum of
inertia of $r$-mode. Moreover, the toroidal magnetic field
$\bar{B}_t$ at the time, $t$, can be calculated as (Cheng \& Yu
2014)
\begin{eqnarray}
\bar{B}_{\rm t}= {15\over8\sqrt{7}\pi}B_{\rm
d}(2\delta+3)\int_0^t\alpha^2\Omega dt'. \label{Bphi}
\end{eqnarray}

Strong toroidal field will subject to the current-driven
instabilities, e.g., Tayler instability. However, the growth rate of
modes of Tayler instability is mainly determined by the toroidal
field configuration, thermal conductivity, stratification, and
rotation, while the latter two effects can lead to the suppression
of the instability (Bonanno \& Urpin 2012; 2013a; 2013b). Meanwhile,
the growth rate of the unstable modes decrease with the increase of
latitude (Bonanno \& Urpin 2012; 2013a; 2013b). If one takes $B_{\rm
t}\propto r^2$ for the toroidal field\footnote{By adopting Equation
(7) of Cuofano \& Drago (2010) for the initial dipole magnetic
field, the toroidal field has the form $B_{\rm t}\propto r^2$ in
most part of the stellar interior. However, in order to simplify the
calculation, the above expression $B_{\rm ini}\propto r^{-3}$ for
initial dipolar field is adopted, thus $B_{\rm t}\propto r^{-1}$. In
fact, the equilibrium strengthes of the amplified toroidal and
poloidal fields are essentially not very sensitive to the initial
dipolar field configuration that one chooses as also suggested in
Cuofano \& Drago (2010).}, at the equator of the star, the growth of
the modes can only be substantially reduced but never be entirely
suppressed in the presence of strong stratification, fast rotation,
and thermal conductivity (Bonanno \& Urpin 2012; 2013a; 2013b). The
reduction of the growth rate of the Tayler's modes will postpone the
occurrence of Tayler instability, leading to the delay of
amplification of the poloidal field. Unfortunately, at present we
cannot quantitatively estimate the delayed timescale and its effect
on the evolutions of the NSs. Following the numerical simulation of
Braithwaite (2006), we qualitatively assume that the Tayler
instability will set in if the condition $\omega_{\rm
A}=\bar{B}_{\rm t}/\left(R\sqrt{4\pi \bar{\rho}}\right)>\Omega$ is
satisfied with $\omega_{\rm A}$ the Alfv\'en frequency. With the
growth of $\bar{B}_{\rm t}$, the NS spins down through
$r$-mode-induced and magnetic deformation-induced GW emissions. This
can naturally result in $\omega_{\rm A}>\Omega$. Generation of a
newly enhanced poloidal field due to the Tayler instability will
close the dynamo loop. Finally, a poloidal-toroidal twisted torus
magnetic field configuration can be formed in the NS interior. When
there is no further accretion on to the NS (after the disruption of
the binary), the enhanced poloidal field could partially diffuse out
to the surface to enhance the exterior dipolar magnetic field
because of Ohmic dissipation and Hall drift. The diffusion timescale
can be roughly estimated as $\tau_{\rm dif}\simeq\tau_{\rm
Hall}\simeq 5\times10^6$ yr, if one takes poloidal field $B_{\rm
pol}=10^{14}$ G, crust thickness $L=1$ km (Goldreich \& Reisenegger
1992). This suggests that if accretion is terminated before the
arising of Tayler instability, the amplification of the surface
dipole field should be delayed by $\sim5\times10^6$ yr with respect
to the occurrence of Tayler instability. Otherwise, the dipole field
will be amplified after $\sim5\times10^6$ yr since the termination
of accretion. The strength of the amplified dipole field is
determined by assuming a simple relation $B_{\rm d}=\xi\bar{B}_{\rm
t}$ (with $\xi =0.01$) (see Cheng \& Yu (2014) for a detailed
discussion).

Finally, the internal temperature $T$ of the accreting NS is
determined by the thermal evolution equation below
\begin{eqnarray}
C_{\rm v}\frac{d T}{d t}=-L_{\nu}-L_{\gamma}+H_{\rm sv}+H_{\rm acc},
\label{deltaT}
\end{eqnarray}
where $C_{\rm v}$, $L_{\nu}$, $L_{\gamma}$, $H_{\rm sv}$, and
$H_{\rm acc}$ are the thermal capacity, neutrino luminosity, surface
photon luminosity, heating due to shear viscous dissipation of
$r$-mode, and accretion heating of the star, respectively. The
expressions for the first four quantities can be found in Yu et al.
(2009), while the last one is taken the same as in Cuofano et al.
(2010).

\section{RESULTS}

\begin{figure}
\centering \resizebox{0.5\hsize}{!}{\includegraphics{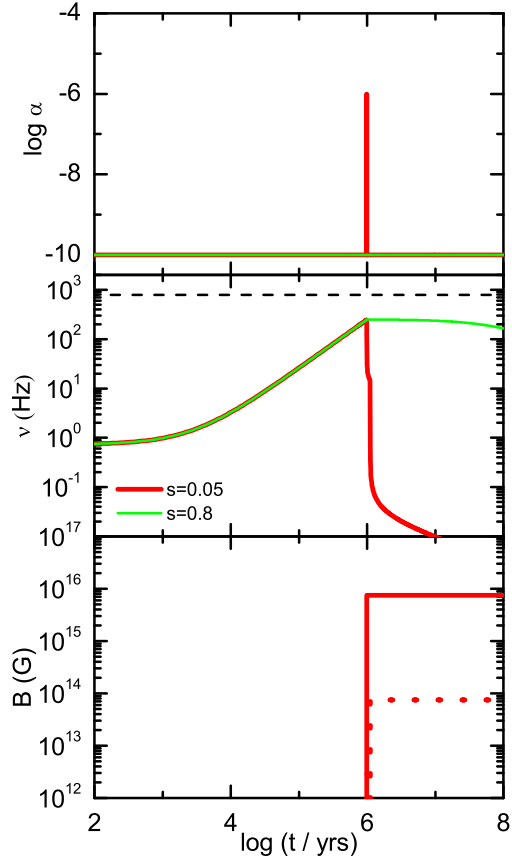}}
   \caption{Evolutions of the $r$-mode amplitudes (top
panel), spin frequencies (middle panel), and strengths of the
magnetic fields (bottom panel) for accreting NSs with different
slippage factors $s$, as shown in the legends. The initial dipolar
magnetic fields of the NSs are taken as $B_{\rm d,i}=10^{9}$ G. The
red solid and dotted lines in the bottom panel represent the
internal toroidal and surface dipolar magnetic fields, respectively.
The spin equilibrium frequency for accreting NSs with $B_{\rm
d,i}=10^{9}$ G, $\dot{M}=6\times10^{-8}M_\odot~{\rm yr}^{-1}$ (black
dashed line) is presented in the middle panel.}
   \label{Fig1}
   \end{figure}

Combining Equations (\ref{alpha}), (\ref{Omega}), (\ref{Bphi}), and
(\ref{deltaT}) we can obtain the evolutions of $r$-modes, spin
frequencies, and magnetic fields of the accreting NSs. The initial
amount of differential rotation $\delta$ is taken the same as in
Cheng \& Yu (2014). Since the results are insensitive to the initial
value of $\alpha$ (Cheng \& Yu 2014), we fix $\alpha_{\rm
i}=10^{-10}$. The initial temperature is taken to be $T_{\rm
i}=10^8$ K, which represents the typical value for accreting NSs
(Cuofano et al. 2012; Haskell et al. 2012). Since we focus on the
evolutions of accreting NSs similar to M82 X-2, we take the present
spin period and magnetic field of M82 X-2 as the initial spin
periods and dipole magnetic fields for these NSs. Thus the initial
spin periods are set as $P_{\rm i}=1.37$ s. While the initial dipole
magnetic fields are taken to be $B_{\rm d,i}=10^8$, and $10^9$ G,
respectively, which are consistent with the strength of M82 X-2, as
suggested in Klu\'zniak \& Lasota (2015). These values may represent
the post-decay strength of the dipole magnetic field. The longest
duration of the early Case B RLOF accretion is assumed to be 1 Myrs
(see Section 1), whose rate is taken as
$\dot{M}=6\times10^{-8}M_\odot~{\rm yr}^{-1}\simeq 3.2\dot{M}_{\rm
Edd}$. Figure \ref{Fig1} shows the evolutions of $\alpha$,
$\nu=(\Omega/2\pi)$, $\bar{B}_{\rm t}$, and $B_{\rm d}$ for the
accreting NSs with $B_{\rm d,i}=10^9$ G, while their $s$ varies.

\begin{figure}
\centering \resizebox{0.5\hsize}{!}{\includegraphics{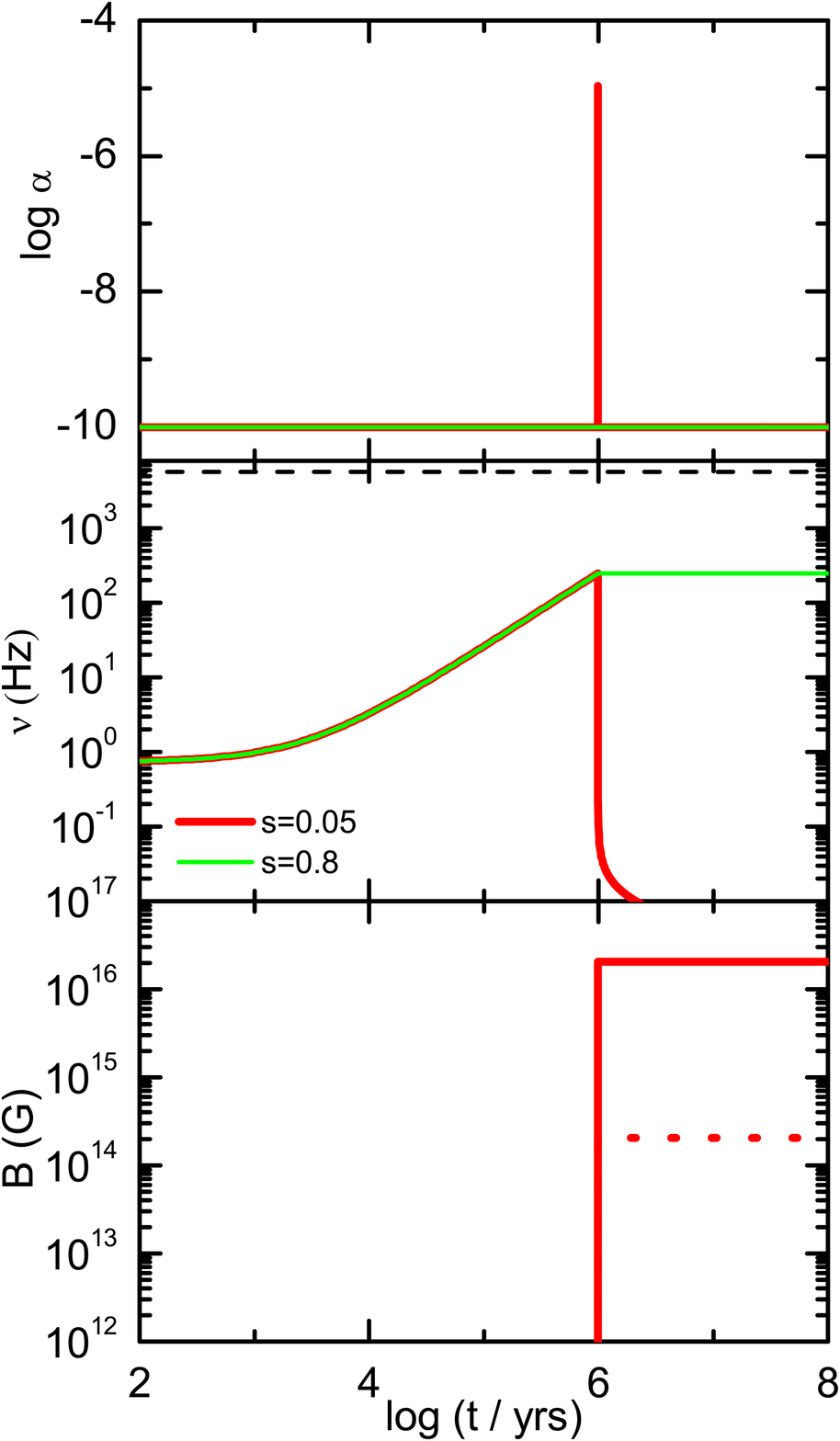}}
\caption{Same as Figure \ref{Fig1}, but $B_{\rm d,i}=10^{8}$ G are
taken. The black dashed line represents the spin equilibrium
frequency for accreting NSs with $B_{\rm d,i}=10^{8}$ G,
$\dot{M}=6\times10^{-8}M_\odot~{\rm yr}^{-1}$.} \label{Fig2}
\end{figure}

As shown in Figure \ref{Fig1}, for the assumed accretion rate and
timescale, the maximum spin frequencies that the accreting NSs with
$B_{\rm d,i}=10^{9}$ G and $P_{\rm i}=1.37$ s could reach are
$\simeq246$ Hz. Similar results were also obtained in Klu\'zniak \&
Lasota (2015), which suggested that M82 X-2 may be spun up to
millisecond periods. However, only the accreting NSs with fairly
small $s(=0.05)$ can reach $\nu_{\rm cr}$ and trigger the $r$-mode
instabilities within 1 Myrs. Since the internal temperature when
thermal equilibrium ($L_{\nu}=H_{\rm acc}$) is achieved can be
determined to be $T\simeq3.9\times10^8$ K, the critical frequency
obtained using Equation (\ref{vcr}) is thus $\nu_{\rm cr}\simeq246$
Hz for $s=0.05$, and 674 Hz for $s=0.8$. Hence, the accreting NSs
with large $s(=0.8)$ will evolve into millisecond rotating NSs
without $r$-modes when RLOF accretion is terminated. It should be
mentioned that the maximum critical frequency is $\simeq 731$ Hz for
$s\simeq 1$, which is still higher than that of the fastest spinning
pulsar observed to date (Hessels et al. 2006). While from Equation
(\ref{veq}), one has $\nu_{\rm eq}(\simeq 784~{\rm Hz})>\nu_{\rm
cr}$ for this set of $B_{\rm d,i}$ and $\dot{M}$. Consequently, the
maximum spin frequency allowed by our model is about 731 Hz, and if
the accretion timescale is long enough, the accreting NSs could
reach this value theoretically. We also note that the spin
evolutions of accreting NSs in IMXBs considered here are different
from those of the NSs in standard high-mass X-ray binaries (HMXBs)
(Urpin et al. 1998). Their various evolution behaviors are mainly
due to (i) different $B_{\rm d,i}$ taken; (ii) different accretion
modes considered (RLOF accretion versus wind accretion/propeller
effect).

As $r$-mode grows and becomes saturated, the toroidal magnetic field
is rapidly amplified to $\sim10^{16}$ G due to $r$-mode-induced
differential rotation. Subsequently, the accreting NS spins down
through $r$-mode and magnetic deformation induced GW emissions.
Meanwhile, with the growth of $\bar{B}_{\rm t}$, the Tayler
instability will arise and then lead to the formation of enhanced
poloidal magnetic field. After the disruption of the NS+evolved star
binary (in post-early Case B mass transfer phase) due to its
frequent encounters with NSs or exchange encounters with other stars
(Sigurdsson \& Phinney 1995; Ivanova et al. 2008; Verbunt \& Freire
2014), accretion on to the NS will be completely terminated. Hence,
the enhanced poloidal field could emerge at the stellar surface and
amplify the surface dipole magnetic field to $\sim10^{14}$ G.
Eventually, the accreting NS could evolve into a magnetar with
$B_{\rm d}\sim10^{14}$ G and $\bar{B}_{\rm t}\sim10^{16}$ G. After
its formation, the magnetar rapidly spins down through
magnetically-induced GW emission and MDR. This magnetar formation
scenario is actually an extrapolation of Cheng \& Yu (2014).
However, there is also one obvious difference: in this work the
magnetars could stem from the accreting NSs with relatively large
$P_{\rm i}$ ($\sim1$ s), whereas in Cheng \& Yu (2014) the magnetars
are produced by newly born isolated fast rotating NSs ($P_{\rm
i}\sim 1$ ms).

In Figure \ref{Fig2}, we show the evolutions of accreting NSs with
$B_{\rm d,i}=10^{8}$ G, while other quantities are kept the same as
in Figure \ref{Fig1}. The lower initial dipole fields do not change
the result qualitatively. In order to trigger $r$-mode and amplify
the toroidal magnetic field within 1 Myrs, $s=0.05$ is required,
which is the same as for $B_{\rm d,i}=10^{9}$ G. The main difference
between $B_{\rm d,i}=10^8$ G and $10^9$ G is, in the former case the
resultant strengthes of $B_{\rm d}$ and $\bar{B}_{\rm t}$ are both
about 3 times higher than those obtained in the latter case. This is
because $r$-mode is less suppressed for lower $B_{\rm d,i}$ (see the
top panels of Figure 1 and 2), thus more energy of the mode can be
converted into magnetic energy. Anyway, for the accreting NS with
$B_{\rm d,i}=10^8$ G, it could also evolve into a magnetar with
$B_{\rm d}\sim10^{14}$ G and $\bar{B}_{\rm t}\sim10^{16}$ G if its
crust is quite elastic. Small $s$ could in fact result in a quite
weak damping effect of $r$-mode, thus a small $\nu_{\rm cr}$, which
can easily be reached through accretion. In contrast, the accreting
NSs with large $s$ could finally evolve into MSPs. We suggest that
the predecessor of SGR 1745-2900 could be an accreting NS in a
special kind of IMXBs, in which the NS was orbiting around an
evolved companion of $M_{\rm c}\sim5.2M_\odot$ with $P_{\rm
orbit}\simeq2.53$ day. Moreover, the accreting NS should have
considerably small slippage factor $s(=0.05)$. Furthermore, the
accreting NSs with $s>0.05$ in such IMXBs could finally evolve into
MSPs.

\section{CONCLUSION AND DISCUSSIONS}

We have investigated the evolutions of accreting NSs in a special
kind of IMXBs, in which the companions are evolved stars in the
Hertzsprung gap phase with $M_{\rm c}\sim5.2M_\odot$, and $P_{\rm
orbit}\simeq2.53$ day. Assuming a duration of 1 Myrs for the early
Case B RLOF accretion, and by adopting a time-averaged rate
$\dot{M}=6\times10^{-8}M_\odot~{\rm yr}^{-1}$, the accreting NSs
with $P_{\rm i}=1.37$ s and $B_{\rm d,i}=10^{8-9}$ G could have
their toroidal and dipolar magnetic fields amplified due to the
$r$-mode-induced differential rotation and Tayler instability. These
accreting NSs could eventually evolve into magnetars with internal
toroidal magnetic fields of $\sim 10^{16}$ G and surface dipole
magnetic fields of $\sim 10^{14}$ G. The precondition is that they
should have quite elastic crusts ($s=0.05$). We note that the
specific evolution process from the accreting NSs to magnetars may
be more complicated, and more exquisite quantitative calculation is
still needed. However, our results suggest the magnetic fields of
some accreting NSs in a special kind of IMXBs could be amplified,
and provide a possible explanation for the origin of the GC magnetar
SGR 1745-2900. Moreover, the accreting NSs with larger $s$ could
finally evolve into MSPs. Hence, both magnetars and MSPs could
originate from accreting NSs in IMXBs. The high stellar density in
the GC should in favor of forming IMXBs, some of which may further
evolve into magnetars and MSPs. We therefore predict more magnetars,
and some of the MSPs should be observed in the GC by future
observations.

The relation between magnetars and NS binaries (both HMXBs and
low-mass X-ray binaries (LMXBs)) is still an open question and has
raised concerns. The possible presence of magnetars in HMXBs has
been proposed in many papers (Bozzo et al. 2008; Doroshenko et al.
2010; Reig et al. 2012; Klus et al. 2014; Tsygankov et al. 2016).
Moreover, it has been suggested that the soft X-ray source 1E
161348-5055 could be a magnetar in a LMXB (Pizzolato et al. 2008).
Our results may also shed light on how magnetars in HMXBs and LMXBs
could be formed. Provided enough angular momentum is available
through accretion from their companions, the accreting NSs with
fairly elastic crusts and relatively small initial spin periods
could eventually have their toroidal magnetic fields amplified. The
accreting NSs would become magnetars, if the enhanced poloidal
fields due to Tayler instability could successfully emerge at the NS
surfaces, and prevent accreted material from falling on the NS
surfaces to avoid field decay. A NS binary comprised of a magnetar
and a companion star may finally be formed. Future research will
focus on the formation of magnetars in such binaries. Furthermore,
in this work we have not considered the interaction between the
magnetic fields and the accretion disk, which can exert an
additional torque on the NS and affect its spin evolution. The
accretion-induced dipole magnetic field decay also has not been
taken into account. We will investigate in detail how these effects
could affect the evolutions of the accreting NSs.

\begin{acknowledgements}
We thank the anonymous referee, X. D. Li, and Y. W. Yu for their
helpful comments and suggestions. This work is supported by the
National Natural Science Foundation of China (grant Nos. 11133002,
and 11178001).
\end{acknowledgements}

\label{lastpage}
\end{document}